\documentstyle[12pt]{article}

\newcommand{\beq}{\begin{equation}}
\newcommand{\eeq}{\end{equation}}
\newcommand{\beqs}{\begin{displaymath}}
\newcommand{\eeqs}{\end{displaymath}}
\newcommand{\beqa}{\begin{eqnarray}}
\newcommand{\eeqa}{\end{eqnarray}}
\newcommand{\beqas}{\begin{eqnarray*}}
\newcommand{\eeqas}{\end{eqnarray*}}

\title{Sturmian Basis Functions for the Harmonic Oscillator}
\author{Frank Antonsen \thanks{e-mail: antonsen@alf.nbi.dk}\\ 
Niels Bohr Institute\\ Blegdamsvej 17, DK-2100
Copenhagen Ø, Denmark}

\begin{document}
\maketitle
\begin{abstract}
We define Sturmian basis functions for the harmonic oscillator and investigate
whether recent insights into Sturmians for Coulomb-like potentials 
can be extended to this important potential. We also treat many body 
problems such as coupling
to a bath of harmonic oscillators. Comments on coupled oscillators
and time-dependent potentials are also made. \\
It is argued that the Sturmian method
amounts to a non-perturbative calculation of the energy levels, but the 
limitations of the method is also pointed out, and the cause of this 
limitation is found to be related to the divergence of the potential. Thus
the divergent nature of the anharmonic potential leads to the Sturmian
method being less acurate than in the Coulomb case. We discuss how modified
anharmonic oscillator potentials, which are well behaved at infinity, leads to
a rapidly converging Sturmian approximation.
\end{abstract}

\section{Introduction}
A typical situation in quantum theory is to be faced with a physical potential
$V$ for which one has to solve the corresponding equation of motion, find
the energies etc. For most realistic potentials it is impossible to find
the energies and wave functions analytically, hence one must resort to
various approximation or numerical schemes. It is the purpose of this paper
to extend one such approximation scheme from Coulomb-like potentials to
harmonic and anharmonic oscillators.\\
The general situation in quantum theory is a system of $N$ particles 
interacting through some potential $V_0(x_1,...,x_N)$. The equation of motion
is then an equation of the general form
\beq
	(D+V_0-E)\psi = 0
\eeq
where $D$ is some differential operator of first or second order, $V_0$ is the
abovementioned potential and $E$ is an eigenvalue. Examples are (units where
$\hbar=c=1$)
\begin{enumerate}
\item The Dirac equation $D=i\gamma^0\gamma^\mu\partial_\mu$ 
(summation over repeated 
indices implied, $\mu=0,1,2,3$, $\gamma^\mu$ are the
Dirac matrices), $V_0=v_0(x)+\gamma^0m$, $v_0(x)$ some potential.
\item The Schr\"{o}dinger equation $D=\frac{1}{2m}\nabla^2$ where 
$\nabla^2$ is the Laplace operator in $d$ dimensions.
\item The Klein-Gordon equation where $D=\Box$, the d'Alembertian operator,
$V_0=m^2+v_0(x)$.
\end{enumerate}
For $N>1$ we simply have $D=\sum_{i=1}^N D_i$ where $D_i$ is the appropriate
operator for the $i$'th particle.
Thus all of relativistic as well as non-relativistic quantum mechanics fall
into this category (even in curved spacetimes). 
Usually one finds a complete set of eigenstates 
corresponding to the different values of $E$, since $D$ is Hermitian these form
an orthonormal set. The problem is, however, that for many cases continuous
as well as discrete eigenvalues have to be taken into account in order
to get a basis for the full Hilbert space. This is for instance the case
in the Coulomb potential case already in non-relativistic quantum
mechanics, where the continuum eigenstates are needed to get completeness.
Consequently, alternatives will have to be found. Shull and L\"{o}wdin, 
\cite{Sturm}, introduced another approach for the
Coulomb potential. Their methods were generalised by Rotenberg who also
coined the word ``Sturmians'' for the new basis set, \cite{Rotenberg}. It has
recently been realised that this approach can be even further generalised
to handle, e.g., many centre potentials, \cite{Avery2} as well as relativistic
effects, \cite{AA} and many-particle systems \cite{Avery4}. 
It is the purpose of this paper to outline the general
theory and to apply it to other potentials such as harmonic oscillators and
variations thereof.\\
Instead of finding a set of eigenfunctions all corresponding to the same
``coupling constants'', i.e., charges for the Coulomb and Yukawa potentials,
$m\omega^2$ for the harmonic oscillator etc., but to different energies as
one normally does, one can take the ``dual'' approach and fix the energy $E$
and then allow the coupling constants to vary. Thus we consider not the
original equation (1) but instead
\beq
	(D+\beta_{\bf n} V_0 - E)\psi_{\bf n} = 0 \label{eq:sturm}
\eeq
where $\beta_{\bf n}$ is some constant depending on the set $\bf n$ of quantum
numbers. The solution of this equation gives $E$ as a function of $\beta_{\bf 
n}$
which can then be inverted to find $\beta_{\bf n}$ as a function of $E$,
assuming we can solve the equation (\ref{eq:sturm}), of course. We will
refer to $\beta_{\bf n}$ as the effective coupling constant -- 
for $V_0=-Z/r$ it
corresponds to scaling the nuclear charge, for $V_0=m\omega^2 x^2$ it 
corresponds to scaling the mass and/or the frequency.\\
The assumed Hermiticity of $D$ then implies
\beq
	(\beta_{\bf n}-\beta_{{\bf n}'})\int \psi_{{\bf n}'}^*V_0\psi_{\bf n} 
	dx = 0 \label{eq:ortho1}
\eeq
i.e., the Sturmian functions satisfy the potential weighted orthogonality
relation
\beq
	\int \psi_{\bf n}^*V_0\psi_{{\bf n}'} dx = N_{\bf n}\delta_{{\bf nn}'}
	\label{eq:ortho2}
\eeq
Strictly speaking (\ref{eq:ortho1}) only implies $\int
\psi_{{\bf n}'}^*V_0\psi_{\bf n}dx = 0$ for 
$\beta_{\bf n}\neq\beta_{{\bf n}'}$.
For sufficiently nice potentials, i.e., potentials without any violent
oscillations, such that the energy $E$ depends monotonically on the quantum
numbers $\bf n$, the orthonormality condition (\ref{eq:ortho2}) follows. An
example of a potential which we do not expect to be able to handle with this
approach is $V_0(x) = \frac{\sin x}{x}$, but potentials $x^{-1},x^{-1}e^{-kx},
x^2$ can be treated this way. It will also be shown later that the many centre
analogues of these potentials also are within reach. Another subtlety concerns
``major'' and ``minor'' quantum numbers in the terminology of Aquilanti and
Avery, \cite{Avery3}. The coefficients $\beta_{\bf n}$ need not depend on
all quantum numbers, those on which it does depend are referred to as ``major''
and the remaining ones are then ``minor''. For the Coulomb potential, for
instance, $\beta_{\bf n}$ only depends on $n$ and not on the angular momentum
quantum numbers $l,m$. The orthogonality with respect to the minor quantum
numbers following from the orthonormality of the spherical harmonics $Y_{lm}$
and the separation of variables, $\psi_{\bf n}(r,\Omega) = \chi_{nl}(r)Y_{lm}
(\Omega)$. For the harmonic oscillator no such subtlety occurs.\\
To solve the Schr\"{o}dinger, Klein-Gordon or Dirac equation for some
physical potential $V$, we begin by considering Sturmians corresponding to
a ``base potential'' $V_0$ for which we can easily solve the corresponding
differential equation to find the basis set. If $V_0$ is sufficiently similar
to $V$, the convergence has been found, for the Coulomb and Yukawa potentials,
 to be very rapid, and, if $E$ is taken
to be the actual physical energy the Sturmians will, furthermore, have 
the right asymptotic behaviour. We will see, however, that this rapid
convergence only takes place when the potential (or rather its matrix 
elements) have sufficiently nice convergence properties themselves.\\
Equation (\ref{eq:ortho2}) has far reaching consequences. Consider a new 
potential $V_0\rightarrow V=
V_0+V'$, where $V_0$ is some potential for which we can easily
find the Sturmians (say, $V_0\sim r^{-1}$ or $V_0\sim x^2$).\footnote{Often,
when working with Sturmians, one would like to take $V'=V$ to be a physical
potential, and only use $V_0$ to find the basis set. We will want to include
$V_0$ as part of the physical potential in this paper, however.}
The
equation of motion for this new system is then
\beq
	(D+V_0+V'-E')\psi = 0
\eeq
and we can expand $\psi$ on the Sturmians for $V$ as $\psi=\sum_{\bf n} 
c_{\bf n}\psi_{\bf n}$
to obtain
\beq
	(D+V_0+V'-E')\sum_{\bf n} c_{\bf n}\psi_{\bf n} = \sum_{\bf n}c_{\bf n}
	\left[(1-\beta_{\bf n})V_0+V'
	+E-E'\right]\psi_{\bf n}
\eeq
upon using (\ref{eq:sturm}). From this we get the secular equation by using
the potential weighted orthogonality relation. Thus
\beq
	\sum_{\bf n}\left[(1-\beta_{\bf n})N_{\bf n}\delta_{{\bf nn}'}+
	\langle\psi_{{\bf n}'}|V'|\psi_{\bf n}\rangle
	+(E-E')\langle\psi_{{\bf n}'}|\psi_{\bf n}\rangle\right] c_{\bf n} = 0
	\label{eq:sec}
\eeq
from which the kinetic energy has disappeared and only the potential $V'$
appears together with the overlap integrals of the Sturmians. It is this
feature of Sturmians which is so important. It often implies that one
can find $E'$ with very great accuracy from an extremely small basis set,
even with just one basis function, $\psi_{\bf n}$, in which case we have
simply
\beqs
	E' = (1-\beta_{\bf n})N_{\bf n}+\langle\psi_{\bf n}|(V'-E)|\psi_{\bf 
	n}\rangle \label{eq:Ep}
\eeqs
giving an explicit formula for $E'$, which when $V'$ is sufficiently close 
to $V_0$ is a surprisingly good fit, \cite{Avery2,AA}.
The reason for this success is to be
found in the very construction of Sturmians. By construction, Sturmians
take the potential much more into account, and thus contain much more
information about the potential. In a sense, Sturmian functions are optimised
with respect to the specific features of the given potential, and it is
precisely this that lies behind their success in the Coulomb case.\\
Another consequence of this can be seen if one attempts to use Sturmians
as a starting point for a variational calculation. Suppose we know the
Sturmians for $V_0$, and now want to use variational theory to estimate the
ground state energy of the Hamiltonian $H=D+V_0+V'$, using the standard
formula
\beqs
	E_0 \leq \frac{\langle\psi|H|\psi\rangle}{\langle\psi|\psi\rangle}
\eeqs
with $\psi=\psi_{\bf n}$, a Sturmian, we get
\beq
	E_0 \leq E+(1+\beta_{\bf n})\frac{\langle\psi_{\bf n}|V_0|
	\psi_{\bf n}\rangle}{\langle
	\psi_{\bf n}|\psi_{\bf n}\rangle}+\frac{\langle\psi_{\bf n}|V'|
	\psi_{\bf n}\rangle}{\langle\psi_{\bf n}|\psi_{\bf n}\rangle}
\eeq
which only concerns the starting potential $V_0$ and the energy $E$ 
to which the Sturmians correspond. If the Hamiltonian, moreover, is of the
form $H=D+V'$, we get
\beq
	E_0 \leq E+\beta_{\bf n}\frac{\langle\psi_{\bf n}|V_0|
	\psi_{\bf n}\rangle}{\langle
	\psi_{\bf n}|\psi_{\bf n}\rangle}+\frac{\langle\psi_{\bf n}|V'|
	\psi_{\bf n}\rangle}{\langle\psi_{\bf n}|\psi_{\bf n}\rangle}
\eeq
which, in any case, is a rather simple integral to compute, suggesting that
Sturmians be a good starting point for variational calculations.\\
It should be noticed, that one will often assume $E=E'$ to be an allowed
energy also of the ``perturbed'' potential $V=V_0+V'$ for an appropriate
value of the effective couplings $\beta_{\bf n}$. In this instance 
(\ref{eq:Ep}) can be seen as showing the physical effect of the 
$\beta_{\bf n}$ as a kind
of screened charge in the Coulomb and Yukawa cases: $(1-\beta_{\bf n})$
measures the ``screening''. Now, $\beta_{\bf n}=1$ corresponds to an unscaled
coupling constant, from $(1-\beta_{\bf n}) \propto \langle \psi_{{\bf n}'}|V'|
\psi_{\bf n}\rangle$ we see that for $E=E'$ to be an allowed energy value
also for $V=V_0$, we must fit the coupling constant $\beta_{\bf n}$ to 
exactly the right value as given by the perturbation $V'$.\\
Since the above consideration only used that $D$ was Hermitian, the results
hold for all the abovementioned cases provided we can find effective couplings
$\beta_{\bf n}$ such that $E$ can be held fixed. It is known that such 
$\beta_{\bf n}$
exist for the Coulomb, Yukawa potentials both in relativistic (the Dirac
equation) and non-relativistic quantum mechanics and also for their many centre
analogues. It will similarly be shown that this also holds for the harmonic
oscillator. In fact, the following argument seems to suggest that it will
always hold. Write the original potential as $V_0(x)=\alpha v(x)$ where 
$\alpha$
is some constant, the original coupling constant, specifying the strength
of the potential, e.g., $\alpha = Z$ for the Coulomb case. Clearly, $E=E
(\alpha)$ if $E$ is an allowed energy eigenvalue. If there is no degeneracy,
a choice for $\alpha$ and for the set of quantum numbers $\bf n$ uniquely
specify the energy $E$, and thus we can invert the relation to find $\alpha$
as a function of $E$ and $\bf n$. If there is degeneracy, the equation $\alpha=
\alpha(E,{\bf n})$ has more than one solution in the physical range, but 
this merely
mean that more than one set of quantum numbers exists giving the same energy,
for each such choice of quantum numbers we get a new solution $\beta_{\bf n}=
\alpha(E,{\bf n})$.\\ 
Another interesting relationship is the momentum space orthonormality relation.
Let $\phi_{\bf n}(k)$ be the Fourier transform of $\psi_{\bf n}(x)$, this then
satisfies
\beq
	(D^t-E)\phi_{\bf n} = -\beta_{\bf n} V_0^t*\phi_{\bf n}
\eeq
where $D^t,V_0^t$ are Fourier transforms of $D,V_0$ and where $*$ denotes 
convolution,
\beq
	V_0^t*\phi_{\bf n} = \int V_0^t(k-k')\phi_{\bf n}(k')dk'
\eeq
Multiply by $\phi_{{\bf n}'}^*$ from the left and perform the $k$-integral
to arrive at
\beq
	\int \phi_{{\bf n}'}^*(D^t-E)\phi_{\bf n} dk = -\beta_{\bf n}
	\int \phi_{{\bf n}'}^* (V_0^t*\phi_{\bf n}) dk
\eeq
Now, by the Fourier convolution theorem, $(fg)^t= f^t*g^t$, and 
Parzival's formula, $\langle f|g\rangle = \langle f^t|g^t\rangle$, we get
\beq
	\int \phi_{{\bf n}'}^*(D^t-E)\phi_{\bf n} dk = -\beta_{\bf n}
	N_{\bf n}\delta_{{\bf nn}'}
\eeq
Hence, the momentum space Sturmians satisfy a weighted orthonormality
relation where the weighting factor is given by the {\em kinetic} part of the
equation of motion and not the {\em potential} as in $x$-space. For the
three operators $D$ mentioned in the beginning, this relation reads
\beqas
	\int \phi_{{\bf n}'}^*(\gamma^0\gamma^\mu k_\mu -E-\gamma^0m)
	\phi_{n} dk &=&
		 -\beta_{\bf n}N_{\bf n}\delta_{{\bf nn}'}\\
	\int \phi_{{\bf n}'}^*\left(k^2-2mE\right)\phi_{\bf n} dk &=&
		-2m\beta_{\bf n}N_{\bf n}\delta_{{\bf nn}'}\\
	\int \phi_{{\bf n}'}^*(k^2-m^2-E)\phi_{\bf n} dk &=&
		-\beta_{\bf n}N_{\bf n}\delta_{{\bf nn}'}
\eeqas
for the Dirac, Schr\"{o}dinger and Klein-Gordon equation respectively. In the
non-relativistic case, one will often write $k_0^2=-2mE$, which is then 
positive ($k_0$ real) for bound states and negative ($k_0$ imaginary) for
unbound states. In this case, the weighting factor becomes $k^2+k_0^2$ which
can be interpreted as the length of the momentum vector in $d+1$ dimensions.
It is this extra dimension which is related to Fock's famous treatment of the
Hydrogen atom where he finds the existence of a $SO(4)$-symmetry. This can be
generalised to arbitrary dimensions by means of hyperspherical harmonics,
\cite{Avery,Avery2}.\\
The momentum space relations have other important implications. The equations
of motion in momentum space is
\beq
	(D^t-E)\phi_{\bf n}(k)=-\beta_{\bf n}V_0^t*\phi_{\bf n}
\eeq
Define, for simplicity, $\tilde{\phi}_{\bf n} = (D^t-E)\phi_{\bf n}$, then we
can write the momentum space orthonormality relation as $\int \phi_{\bf n}^*
\tilde{\phi}_{{\bf n}'} dk=-\beta_{\bf n}N_{\bf n}\delta_{{\bf nn}'}$ and the
equation of motion as
\beqa
	\tilde{\phi}_{\bf n}(k) &=& -\beta_{\bf n}(V_0^t*\phi_{\bf n})(k) \\
	&=& -\beta_{\bf n}\int V_0^t(k-k')\phi_{\bf n}(k')dk' \label{eq:keq}
\eeqa
Now, make the following {\em Ansatz}
\beq
	V^t_0(k-k') = \sum_{\bf n} c_{\bf n}\tilde{\phi}_{\bf n}(k)
	\tilde{\phi}^*_{\bf n}(k')
\eeq
Inserting this into (\ref{eq:keq}) we then get
\beq
	c_{\bf n} = \frac{N_{\bf n}}{\beta_{\bf n}^2}
\eeq
i.e.,
\beqa
	V_0^t(k-k') &=& \sum_{\bf n}\frac{N_{\bf n}}{\beta_{\bf n}^2}
	\tilde{\phi}_{\bf n}(k)\tilde{\phi}_{\bf n}^*(k')\\
	&:=& \sum_{\bf n}\frac{N_{\bf n}}{\beta_{\bf n}^2} D^t(k)(D^t(k'))^*
	\phi_{\bf n}(k)\phi_{\bf n}^*(k')
\eeqa
When $V_0^t$ is some function of $k,k'$, this is a useful summation formula
for the momentum space Sturmians. When $V^t_0$ is a differential operator,
on the other hand, as happens when $V_0=x^\gamma,\gamma>0$, then this is
a spectral representation of that operator.\\
It should be emphasised that the coefficients in the above expansion are
very simple in the sense that they are the natural quantities related to the
basis set, namely the normalisation, the ``effective charge'' $\beta_{\bf n}$,
and the kinetic operator (which is then a polynomial in momentum space).
This shows more precisely how Sturmians are adapted to the potential.\\
We will now turn to the specific case of a harmonic oscillator.

\section{The Harmonic Oscillator}
To begin with we work with $N=1$ and in $d=1$ dimension. The potential is
$V_0=\frac{1}{2}x^2$ and we solve the equation
\beq
	\left(-\frac{1}{2m}\frac{d^2}{dx^2}+\frac{1}{2}\beta_{\bf n} x^2-
	E\right)\psi_{\bf n}=0
\eeq
which is the harmonic oscillator Schr\"{o}dinger equation with $m\omega^2$
replaced by $\beta_{\bf n}$. We will henceforth use mass weighted coordinates
and put $m=1$, if the mass needs to be reinstated one simply replaces $x$ by
$\sqrt{m}x$. In this case $\omega$ in the original harmonic oscillator
Hamiltonian has been replaced by $\beta_{\bf n}^{1/2}$.
The solution of this is clearly
\beq
	\psi_{\bf n}(x) = \pi^{-1/4} (n!)^{-1/2} 2^{-n/2}H_n((\beta_{\bf n} 
	)^{1/4}x)e^{-\frac{1}{2}\beta_{\bf n}^{1/2}x^2}
\eeq
where ${\bf n} = n$ is a non-negative integer, $H_n$ is a Hermite polynomial 
and $E=\omega(n+\frac{1}{2})=\sqrt{\beta_{\bf n}}(n+\frac{1}{2})$. 
From this relationship between $E,{\bf n},\beta_{\bf n}$ we read off
\beq
	\beta_{\bf n} = \left(\frac{E}{n+\frac{1}{2}}\right)^2
\eeq
which is the promised relationship between the effective coupling, the
quantum number and the energy, needed to make the Sturmian machinery work.\\
The effect of scaling the argument by an $n$-dependent quantity is to scale
all the functions to take values within the same interval. This is illustrated
in figure 1.\\
The orthonormality relation, $\langle\psi_{\bf n}|\frac{1}{2}x^2|\psi_{\bf n'}
\rangle = N_{\bf n}\delta_{\bf nn'}$, then reads
\beq
	\int H_n(\beta_{\bf n}^{1/4} x)H_{n'}((\beta_{{\bf n}'})^{1/4}x)
	e^{-\frac{1}{2}(\beta_{\bf n}^{1/2}+\beta_{{\bf n}'}^{1/2})x^2}
	x^2 dx = N_{\bf n} \sqrt{\pi} 2^{n+1} n! \delta_{{\bf nn}'}
\eeq
The first few normalisation factors $N_{\bf n}$ turn out to be as in table 1,
we see that $N_{\bf n} \propto E^{-3/2}$ -- in general, $\langle\psi_{\bf n}|
x^k|\psi_{\bf n'}\rangle\propto E^{-(k+1)/2}$ for $k=0,1,2,...$ This is easily
seen from a simple scaling argument: Perform the scaling $x\rightarrow 
\beta_{\bf n}^{1/4}x\sim
E^{1/2}x$ (for the sake of this argument we can ignore the $n$ dependence), in
the integral and the result follows directly.\\
From table 1 it is clear that
\beq
	N_n=\frac{(2n+1)^{3/2}}{8\sqrt{2}} E^{-3/2} = \frac{1}{4}
	\beta_{\bf n}^{-3/4}
\eeq
Notice that the Sturmian orthonormality relation above differs slightly 
from the usual one in
two ways, (1) the Hermite polynomials have different arguments (normally
the argument is just $\sqrt{\omega}x$, irrespective of the value of the
quantum number $n$), and (2) the appearance of the factor
$x^2$.\\
Similarly, the overlap matrix between the first five Sturmians $T_{{\bf nn}'}
=\int\psi_{\bf n}^*\psi_{{\bf n}'}dx$ is
\beq
	T=E^{-1/2}\left(\begin{array}{ccccc}
	\frac{1}{\sqrt{2}} & 0 & -\frac{1}{3}\sqrt{\frac{5}{3}} & 0 & 
		\frac{12}{25}\sqrt{\frac{3}{5}}\\
	0 & \sqrt{\frac{3}{2}} & 0 & -\frac{21}{25}\sqrt{\frac{3}{5}} & 0\\
	-\frac{1}{3}\sqrt{\frac{5}{3}} & 0 & \sqrt{\frac{5}{2}} & 0 &
		-\frac{132}{343}\sqrt{\frac{30}{7}}\\
	0 & -\frac{21}{25}\sqrt{\frac{3}{5}} & 0 & \sqrt{\frac{7}{2}} &0\\
	\frac{12}{25}\sqrt{\frac{3}{5}} & 0 & -\frac{132}{343}	
		\sqrt{\frac{30}{7}} & 0 & \frac{3}{\sqrt{2}}
	\end{array}\right)
\eeq
For $d>1$, the energy spectrum is $E=\omega (n+d/2)$ leading to a very simple
modification in the expression linking $\beta_{\bf n}$ and $E$. The wave
functions become products of Hermite polynomials and ${\bf n}=(n_1,...,n_d), n=
n_1+n_2+...+n_d$
\beq
	\psi_{\bf n} = \pi^{-d/4}2^{-dn/2}\left(\prod_{i=1}^d(n_i!)^{-1/2}
	H_{n_i}(\beta_{\bf n}^{1/4}x_i)\right)
	e^{-\frac{1}{2}\beta_{\bf n}^{1/2} (x_1^2+...x_d^2)}
\eeq
Note, $\beta_{\bf n} = \beta_n$ depends only on $n=n_1+...n_d$, the ``total
quantum number''. The case of more than one particle merely corresponds to
a harmonic oscillator in $D=dN$ dimensions, where $N$ denotes the number of
particles and $d$ the number of spatial dimensions. If the particles have
different masses, one have to use mass-weighted coordinates, $x_i\mapsto
y_i=\sqrt{m_i}x_i$, but otherwise no modifications are needed. Clearly, all
the important features can be found already in the $d=N=1$ case, for which
reason we will stick to this situation in the following unless otherwise 
stated.\\
The momentum space formulation of the harmonic oscillator needs the Fourier
transform of the Hermite polynomials. It is proven in the appendix that
\beqa
	\phi_{\bf n}(k)&=& \pi^{-1/4}(n!)^{-1/2}2^{-n/2}
	\sqrt{\frac{2\pi}{\beta_{\bf n}^{1/4}}}\left(1+\frac{1}{2}(\beta_
	{\bf n})^{1/4}\right)\times\nonumber\\
	&&H_n(ik(1+\frac{1}{2}\beta_{\bf n}^{1/4})^{-
	1/2})e^{\frac{1}{2}k^2 \beta_{\bf n}^{-1/2}}
\eeqa
is the Fourier transform of $\psi_{\bf n}$. The Hermite polynomials of
imaginary arguments appear in this formula, this means that for $n$ even
$\phi_{\bf n}$ is purely real, whereas for $n$ odd it is purely imaginary.
The momentum space orthogonality relation then gives the following new
relationship between Hermite polynomials
\beqa
	&&\int H_n(ik(1+\frac{1}{2}\beta_{\bf n}^{1/4})^{-1/2}) H_{n'}
	(ik(1+\frac{1}{2}(\beta_{{\bf n}'})^{1/4})^{-1/2}) e^{\frac{1}{2}k^2
	(\beta_{\bf n}^{1/2}+\beta_{{\bf n}'}^{1/2})}
	(k^2-2E) dk\nonumber\\
	&&\qquad = -\frac{1}{\sqrt{2\pi}} n! 2^{n+1}\beta_{\bf 
	n}^{5/4}N_{\bf n} \left(1+\frac{1}{2}\beta_{\bf n}^{1/4}
	\right)^{-2}\delta_{{\bf nn}'}
\eeqa
a somewhat unexpected result.

\section{The Anharmonic Oscillator}
We now add a new potential $V'=\alpha x^3$ to $V$. We then need to compute
the matrix elements of this in the basis of harmonic oscillator Sturmians.
We need to compute the matrix $W^{(3)}_{{\bf nn}'}=\int\psi_{\bf n}^* x^3
\psi_{{\bf n}'}dx$. For ${\bf n,n}'=0,1,...,4$ we get the following explicit
result
\beq
	W^{(3)}=E^{-2}\left(\begin{array}{ccccc} 
	0 & \frac{27}{64} & 0 & -\frac{343\sqrt{3}}{2048} & 0\\
	\frac{27}{64} & 0 & \frac{7425\sqrt{5}}{2048} & 0 
		&-\frac{3159}{512\sqrt{2}}\\
	0 & \frac{7425\sqrt{5}}{2048} & 0 &\frac{677425\sqrt{5}}{41472} &0\\
	-\frac{343\sqrt{3}}{2048} & 0 &\frac{677425\sqrt{5}}{41472} & 0 &
		\frac{431831169}{4194304}\\
	0 & -\frac{3159}{512\sqrt{2}} & 0 & \frac{431831169}{4194304} &0
	\end{array}\right)
\eeq
The secular equation then reads
\beq
	\det((1-\beta_{\bf n})N_{\bf n}\delta_{{\bf nn}'}+\alpha W^{(3)})=0
\eeq
By including the first $N$ Sturmians we get the ground state
energies shown in table 2 with $\alpha=.1$.\\
\noindent Another version is $V'= x^4$, for this case the matrix 
$W^{(4)}_{\bf nn'}:=\int\psi_{\bf n}^*x^4\psi_{\bf n'}dx$ becomes
\beq
	W^{(4)}= E^{-5/2}\left(\begin{array}{ccccc}
	\frac{3}{16\sqrt{2}} & 0 & \frac{25}{144}\sqrt{\frac{5}{3}} & 0 &
		-\frac{243}{1000}\sqrt{\frac{3}{5}}\\
	0 & \frac{135}{16}\sqrt{\frac{3}{2}} & 0 & \frac{27783}{2000}
		\sqrt{\frac{3}{5}} & 0\\
	\frac{25}{144}\sqrt{\frac{5}{3}} & 0 &\frac{975}{16}\sqrt{\frac{5}{2}}
		& 0 &\frac{625725}{9604}\sqrt{\frac{15}{14}}\\
	0 & \frac{27783}{2000}\sqrt{\frac{3}{5}} & 0 & \frac{3675}{16}
		\sqrt{\frac{7}{2}} & 0\\
	-\frac{243}{1000}\sqrt{\frac{3}{5}} & 0 & \frac{625725}{9604}
		\sqrt{\frac{15}{14}} & 0 & \frac{29889}{16\sqrt{2}}
	\end{array}\right)
\eeq
As for $x^3$ we get the energies $E$ by including the first $N=1,...,5$
Sturmians also shown in table 2. As is apparent from the table, the ground
state values are unstable, i.e., the approximation has failed to converge.
This is in sharp contrast to what is known to be the case for the Coulomb and
Yukawa potentials (both in their single as in their many-center form) where the
convergence is very rapid. This failure can be traced back to the 
non-convergence of the matrix elements $W_{\bf nn'}^{(k)}\rightarrow\infty$
for $n,n'\rightarrow \infty$, which again is a consequence of the divergent
behaviour of the potentials as $x\rightarrow\infty$. Although the Coulomb and
Yukawa potentials are singular at the origin, $r=0$, their matrix elements
none the less fall of rather rapidly as the quantum numbers increase, due to
the quick fall-off of the potentials themselves as $r$ increases.\\
Standard perturbation theory would give a value for the ground state energy
in the two cases of
\beqa
	E_0' &=& E_0-\frac{11}{8}\frac{\hbar^2\alpha^2}{m^3\omega^4}
	\qquad \mbox{(cubic potential)}\\
	E_0' &=& E_0+\frac{3}{16}\frac{\hbar^2\alpha}{m^2\omega^2}-
	\frac{23}{4}\frac{\hbar\alpha^2}{m^2\omega^3}\qquad 
\mbox{(quartic potential)}
\eeqa
by going to second order in the coupling constant $\epsilon$ and using standard
Rayleigh-Schr\"{o}dinger perturbation theory, which gives a non-convergent
series -- for the particular example of $m=\hbar=\omega=1, \alpha=.1$ we
get $E_0'=0.3625$ for the cubic and $E_0'=0.46125$ for the quartic anharmonic
oscillator. Clearly, this is not in good agreement with the result found by
using Sturmians, but as will be seen below, this is due to the divergence of 
the perturbation series. That the Sturmian method is non-perturbative is
suggested by the general solution of the secular equation, for the quartic
anharmonic potential the ground state energy as a function of $\alpha$ is
found to be (for $N=1$) $E_0'=z^{1/3}+\frac{1}{12}z^{-1/3}$ where
\beqs
	z=\frac{2}{2592\alpha+\sqrt{6718464\alpha^2-6912}}
\eeqs
Such a dependence of $E$ on $\alpha$ can only be obtained in perturbation
theory by performing at least a partial resummation of the infinite series.
That the perturbation series is divergent is mirrored in the behaviour of
$E_0'$ as a function of the coupling constant $\alpha$: As $\alpha$ increases
so does the real part of $E_0'$. Furthermore, $E_0'$ has a small imaginary
part which is wildly oscillating but vanishes for $\alpha > 0.003$ and hence
quickly becomes unimportant.\\
Thus, we must compare our Sturmian energies with non-perturbative results.
The divergence of the perturbation series can be improved in a number of
ways, as done recently by Bender and Bettencourt, \cite{BB}, and by
Kunihiro, \cite{Kunihiro}. The latter paper performs a re-summation of the
perturbation series by using a renormalisation group (RG) improved
technique. From this very rapid expressions for the ground state energy for
the quartic anharmonic oscillator is found. Some high-precision numerical 
results for the quartic anharmonic oscillator have been found by Bacus et al.,
\cite{Bacus}, and we will make a comparison with those findings. Other recent
papers on related topics are \cite{others}.\\
We will go back to the original secular equation, (\ref{eq:sec}), and let
$E$ be the energy of the harmonic oscillator. When $E=E_n=(n+1/2)\hbar\omega$ 
we get $\beta_{\bf n}=1$ (we will say we are ``on shell'') and the secular
equation simplifies, since the results then no-longer depend on the
normalisation factor $N_{\bf n}\propto E^{-3/2}$.
 Let the number of Sturmians in our basis set be $N$, then
for $N=1$ we get simply
\beq
	E'=E+\alpha \frac{\langle\psi_{\bf n}|x^4|\psi_{\bf n}\rangle}{
	\langle\psi_{\bf n}|\psi_{\bf n}\rangle} \equiv
	E+\alpha T_{\bf nn}^{-1}W_{\bf nn}^{(4)}
\eeq
Let $N=2$ and let the basis set correspond
to the quantum numbers $n,n+1$, we then find $E'$ by solving the quadratic
equation
\beqa
	0&=&(E-E')^2 T_{\bf nn}T_{\bf n+1,n+1}+\alpha (E-E')(T_{\bf nn}
	W_{\bf n+1,n+1}^{(4)}+\nonumber\\
	&&\qquad T_{\bf n+1,n+1}W_{\bf nn}^{(4)})+
	\alpha^2W_{\bf nn}^{(4)}W_{\bf n+1,n+1}^{(4)}
\eeqa
For $n=0,1,..,4$ we get the results shown in the table 3, where we also show
the highly accurate (approx. 90 significant digits) of Bacus et al.. We note
that this crude Sturmian approximation is able to get the right order of 
magnitude for the energy states even of the excited states, but we also 
notice that the accuracy decreases as $n$ increases. This is once more due to 
the non-convergence of the matrix elements of the pertubation potential.
Increasing the basis set will not lead to improved accuracy, but can in fact
lead to quite the opposite because of this divergence. Hence, the best results
are found by the simplest approximation, namely $N=1$. In contrast to the
case of asymptotically well-behaved potentials such as the Coulomb and Yukawa
ones, hitherto studied in the literature.

\section{A Comment on Time Dependent Potentials: The Damped Oscillator}
Another important variant of the harmonic oscillator is the damped oscillator.
The particular Hamiltonian we're going to study is the so-called 
Caldirola-Kanai oscillator, \cite{damped},
\beq
	H=\frac{p^2}{2}e^{-2\gamma t} + \frac{1}{2}\omega x^2 e^{2\gamma t}
\eeq
where $\gamma$ is some constant, the friction the coefficient and $t$ denotes 
time. This can be re-expressed in terms of a potential $V'$
\beq
	V'=\frac{1}{2}\omega^2 x^2e^{2\gamma t'(t)} \equiv \xi(t)V_0(x)
\eeq
with
\beq
	t'(t)=\frac{1-e^{-2\gamma t}}{\gamma}
\eeq
For time dependent potentials $V'$, the secular equation have to be modified.
The Schr\"{o}dinger equation for the full system reads
\beq
	(D+V_0+V'(t))\psi= i\hbar \frac{\partial}{\partial t}\psi
\eeq
expanding $\psi=\sum_{\bf n}c_{\bf n}(t)\psi_{\bf n}$ leads to the modified
secular equation
\beq
	\sum_{\bf n}\left[(1-\beta_{\bf n})N_{\bf n}\delta_{{\bf nn}'}+
	\langle\psi_{{\bf n}'}|V'(t)|\psi_{\bf n}\rangle+E\langle
	\psi_{{\bf n}'}
	|\psi_{\bf n}\rangle\right]c_{\bf n} = i\hbar\sum_{\bf n}\dot{c}_{\bf 
	n}\langle\psi_{{\bf n}'}|\psi_{\bf n}\rangle
\eeq
Using the relationship between $V'$ and $V_0$ we can rewrite this as
\beq
	\sum_{\bf n}\left[(1-\beta_{\bf n}(1+\xi(t)))N_{\bf n}\delta_{{\bf 
	nn}'}+E\langle\psi_{{\bf n}'}|\psi_{\bf n}\rangle\right] c_{\bf n}
	=i\hbar\sum_{\bf n}\dot{c}_{\bf n}\langle\psi_{{\bf n}'}|\psi_{\bf n}
	\rangle
\eeq
For the simplest possible case of only including $N=1$ Sturmians in the basis
set, the solution to this secular equation is of course
\beq
	c_{\bf n}(t)=c_{\bf n}(0)\exp\left(-it\left[(1-\beta_{\bf n})
	\frac{N_{\bf n}}{T_{\bf nn}}+E\right]-
	i(1-\beta_{\bf n})\frac{N_{\bf n}}{T_{\bf nn}}\int_0^t\xi(t')dt'\right)
\eeq
It turns out that one can actually compute the integral, since
\beqs
	\xi(t)=e^{2-2e^{-2\gamma t}}
\eeqs
and the integral can be expressed in terms of the exponential integral function
leading to
\beq
	c_{\bf n}(t)= c_{\bf n}(0) e^{-i\omega_0 t-i\omega(t)}
\eeq
where
\beqa
	\omega_0 &=& (1-\beta_{\bf n})\frac{N_{\bf n}}{T_{\bf nn}}+E\\
	\omega(t) &=& (1-\beta_{\bf n})\frac{N_{\bf n}}{T_{\bf nn}}
	\frac{e^2}{2\gamma}\left(
	{\rm Ei}(-2)-{\rm Ei}(-2e^{-2\gamma t})\right)
\eeqa
Naturally, ``on shell'' (i.e., for $\beta_{\bf n}=1$) we simply get $c_{\bf n}
(t)=c_{\bf n}(0) e^{-iEt}$ as one would expect, whereas ``off 
shell'' (i.e., for $\beta_{\bf n}\neq 1$) we get a highly oscillatory 
behaviour. The explicit results for $N=1,n=0$ are $\omega_0=\frac{1+4E^2}{8E},
\omega(t)=\frac{e^2(1-4E^2)}{16 E\gamma}\left({\rm Ei}(-2)-{\rm Ei}(-2
e^{-2\gamma t})\right)$.
One should also notice that this result holds even for $N\neq 1$,
one must then interpret the division by $T_{\bf nn}$ as multiplication from
the left by the inverse $T_{\bf nn'}^{-1}$, whereby $\omega_0,\omega(t)$
become matrices.\\
The real and imaginary parts of $c_0(t)\psi_0(x)$ have been plotted in figure
2a-b. We notice that the oscillations in the $t$-direction die out either
as $x$ increases (because of the decrease of $\psi_0(x)$) or as time goes.\\
This simple example shows how time-dependent problems simplify in the
Sturmian approach because of the simplification of the secular equation. Thus,
Sturmians are well suited for problems with time-dependent potentials or for
scattering processes. But they are of course subject to the same limitations
as in the time-independent case.\\
It also shows that this particular time-dependent damped oscillator is exactly
solvable using the Sturmian approach.

\section{A Bath of Harmonic Oscillators}
Consider the potential
\beq
	V=\frac{1}{2}\sum_i g_i (x-x_i)^2
\eeq
where $i$ runs over some index set. This represents the potential coming from
a family, indexed by $i$, of harmonic oscillators situated at $x_i$ and with
coupling constants (characteristic frequencies) $g_i$. We will usually restrict
ourselves to $i$ being discrete corresponding to an infinite lattice of
oscillators, but for field theoretical purposes it can also be relevant to
allow $i$ to run over a continuous index set (in which case the sum must be
interpreted as an integral). This example is the harmonic oscillator
analogue of the many centre Coulomb potential treated in \cite{Avery2,Avery3}.
Note that we can rewrite $V$ as
\beqa
	V&=&\frac{1}{2}\bar{g}x^2 - x\sum_i g_ix_i +\frac{1}{2}\sum_ig_ix_i^2\\
	&\equiv & \bar{g}V_0-x c_1+ c_2
\eeqa
where $\bar{g}=\sum_i g_i, c_1=\sum_ig_ix_i, c_2=\frac{1}{2}\sum_ig_ix_i^2$, 
hence we can see this as a perturbation of the
original potential $V_0$. Because of this feature of the bath of harmonic
oscillators, our computations will simplify somewhat from the many-centre
Sturmians introduced for the Coulomb potential in \cite{Avery2,Avery3}, which
is very fortunate since those papers use a Fourier transform approach which is
not useful for the harmonic oscillator -- as mentioned earlier, the Fourier
transform, $V^t(k)$, of the potential $V$ is a differential operator for
the harmonic oscillator (actually, the Hamiltonian is form-invariant under
Fourier transforms).\\
Furthermore, we can complete the squares to write $V$ as another harmonic 
oscillator plus a constant, in fact
\beq
	V=\frac{1}{2}\bar{g}\left(x-\frac{c_1}{2\bar{g}}\right)^2+c_2-
	\frac{c_1^2}{4\bar{g}}
\eeq
Consequently, we can obtain a solution to the many-centre Schr\"{o}dinger
equation by simply making the following substitutions in the solution for the
single harmonic oscillator
\beq
	x \rightarrow x+\frac{c_1}{2\bar{g}}\qquad \beta_{\bf n}\rightarrow
	\bar{g}\beta_{\bf n}\qquad E\rightarrow E-c_2+\frac{c_1^2}{4\bar{g}}
\eeq
i.e., the new $\beta_{\bf n}$, $\tilde{\beta}_{\bf n}$ reads
\beqs
	\tilde{\beta}_{\bf n} = \bar{g}\left(\frac{E-c_2+c_1^2/4\bar{g}}{n+1/2}
	\right)^2
\eeqs
Hence the Sturmians read
\beq
	\Psi_{\bf n}(x;x_i) = \pi^{-1/4}(n!)^{-1/2}2^{-n/2}H_n\left(
	\tilde{\beta}_{\bf n}^{1/4}(x+\frac{c_1^2}{4\bar{g}})\right)
	e^{-\frac{1}{2}\sqrt{\tilde{\beta}_{\bf n}}
	(x+\frac{c_1^2}{4\bar{g}})^2}
\eeq
These are then a convenient basis for many-centre problems.\\
For the many-centre Coulomb potential, the Sturmians become related to the
one-centre Sturmians by means of phase factors $e^{ik_ix_i}$, whereas for
the harmonic oscillators the many-centre Sturmians are related to the
one-centre ones by means of a translation $x\rightarrow x-\frac{c_1^2}
{4\bar{g}}$ as well as a scaling and a shift in energy.\\
As a simple example, consider a particle moving in a potential coming from
harmonic oscillators situated at $x_i=i, i=1,...,M$ all with equal strengths,
$g_i=1$. Then $\bar{g}=M, c_1=\frac{1}{2}M(M+1), c_2=\frac{1}{12}M(M+1)(2M+1)$
and
\beqs
	x\rightarrow x+\frac{1}{4}(M+1) \qquad E\rightarrow E-\frac{1}{48}
	M(5M^2+6M+1)
\eeqs
thus
\beqs
	\tilde{\beta}_{\bf n} = M\left(\frac{96 E- 2M(5M^2+6M+1)}{48 (2n+1)}
	\right)^2
\eeqs
and the first two Sturmians read explicitly
\beqa
	\Psi_0(x;x_i) &=& \pi^{-1/4} \exp\left(-M\frac{48E- M(5M^2+6M+1)}{48}
	\left(x+\frac{1}{4}(M+1)\right)^2\right)\\
	\Psi_1(x;x_i) &=& \pi^{-1/4} \frac{1}{6}\sqrt{M(48 E- M(5M^2+6M+1))}
	\left(x+\frac{1}{4}(M+1)\right) \times
	\nonumber\\
	&&\qquad \exp\left(-M\frac{48 E- M(5M^2+6M+1)}
	{144}\left(x+\frac{1}{4}(M+1)\right)^2\right)
\eeqa
The energy is found from the Schr\"{o}dinger equation which leads to the ``on
shell'' condition $\tilde{\beta}_n=1$. Consequently, the energy of the $n$'th
state is
\beq
	E_n=(n+\frac{1}{2})\bar{g}^{-1/2}-c_2+\frac{c_1^2}{4\bar{g}}
\eeq
irrespective of the number of Sturmians used as the secular equation (by
construction) is diagonal. In the particularly simple case of $M$ evenly
spaced oscillators all with the same value of the coupling $g_i=1$ 
reads\footnote{Note, for $M=1$, $E\neq n+1/2$, because the potential in this
instance is $V=\frac{1}{2}(x-1)^2$ and {\em not} $V=V_0$. Had we instead used
$x_i=i-1$, we would get $V(M=1)=V_0, E(M=1)=n+1/2$. In that case, by the
way, $c_1=\frac{1}{2}M(M-1), c_2=\frac{1}{12}M(M-1)(2M-1)$.}
\beq
	E_n=(n+\frac{1}{2})M^{-1/2}-\frac{1}{12}M (M+1)(2M+1)+\frac{1}{16}
	M(M+1)^2
\eeq
which is then the energy of a (non relativistic) particle moving in a 
one-dimensional lattice of oscillators -- a highly simplified model of, say,
a particle in a solid. For a many-dimensional lattice we would simply use
products of unidiemensional Sturmians.

\section{A Comment on Coupled Oscillators}
Consider now a potential of the form
\beq
	V= \frac{1}{2}\sum_i g_i(x-x_i)^2+\frac{1}{2}\sum_{i\neq j}\lambda_{ij}
	(x-x_i)^2(x-x_j)^2
\eeq
which introduces a coupling between the oscillators at the various positions.
In a manner similar to the manipulations of the bath of oscillators, this can
be transformed into a single anharmonic oscillator potential
\beq
	V=\frac{1}{2}\bar{g}x^2-c_1x+c_2-c_3x^3+c_4x^4
\eeq
where
\beqa
	\bar{g} &=&\sum_ig_i\\
	c_1&=&\sum_i 2(g_i-\sum_{j\neq i}\lambda_{ij}(x_j^2+x_jx_i))x_i\\
	c_2&=& \frac{1}{2}\sum_i(g_i+\sum_{j\neq i}x_j^2)x_i^2\\
	c_3&=&\sum_{i\neq j}\lambda_{ij}(x_i+x_j)\\
	c_4&=&\frac{1}{2}\sum_{i\neq j}\lambda_{ij}
\eeqa
Suppose $\chi$ is a solution to the corresponding Schr\"{o}dinger equation.
We can then expand $\chi$ either on the ordinary Sturmians, $\psi_{\bf n}$,
or the Sturmians for a bath of oscillators, $\Psi_{\bf n}$. If we choose the
latter option, we have to complete the squares to obtain the centre of the
new oscillator, but this would mean that the anharmonic terms too would have
to be shifted and this would again introduce lower powers of the new, shifted
position. Hence the secular equation would end up having the same structure
and thus the same level of complication. Consequently, nothing is lost by
expanding on the 
single oscillator Sturmians $\psi_{\bf n}$, $\chi=\sum_{\bf n}\alpha_{\bf n}
\psi_{\bf n}$. The secular equation then becomes
\beq
	0=\sum_{\bf n'}\left[(\beta_{\bf n}+\bar{g})N_{\bf n}\delta_{\bf nn'}
	-c_1W^{(1)}_{\bf nn'}+c_2T_{\bf nn'}-c_3W^{(3)}_{\bf nn'}
	-c_4W^{(4)}_{\bf nn'}\right] \alpha_{\bf n'}
\eeq
where
\beq
	W_{\bf nn'}^{(k)} := \int \psi_{\bf n}^*x^k\psi_{\bf n'} dx
\eeq
is the matrix elements of the $k$'th power of $x$, $W^{(0)}_{\bf nn'}=
T_{\bf nn'}$. The only one of these we do not already know is for $k=1$
in which instance a straightforward computation yields
\beq
	W^{(1)} = E^{-1}\left(\begin{array}{ccccc}
		0 & \frac{3}{8} & 0 & -\frac{21}{128}\sqrt{3} & 0\\
		\frac{3}{8} & 0 & \frac{75}{128}\sqrt{3} & 0 & 
			-\frac{81}{64\sqrt{2}}\\
		0 & \frac{75}{128}\sqrt{3} & 0 & \frac{1925}{1728}\sqrt{5}
			& 0\\
		-\frac{21}{128}\sqrt{3} & 0 & \frac{1925}{1728}\sqrt{5} & 0 &
			\frac{508599}{131072}\\
		0 & -\frac{81}{64\sqrt{2}} & 0 &\frac{508599}{131072} & 0
	\end{array}\right)
\eeq
for the first five Sturmians. Consequently, for $N=1$ we get (since 
$W^{(2k+1)}$ is diagonal)
\beq
	(\beta_{\bf n}+\bar{g})N_{\bf n}+c_2T_{\bf nn}-c_4W^{(4)}_{\bf nn}=0
\eeq
which is a cubic equation for $E$, e.g. for $n=0$
\beqs
	\bar{g}-3 c_4 E+4 E^2 + 16 c_2 E^3 =0
\eeqs
which for the extremely simple case of $\bar{g}=c_2=c_4=1$ has the three 
solutions $E=-0.669498, 0.209749\pm i 0.222168$, i.e., one negative energy
state (hence a bound state) and two complex conjugate oscillatory states.\\
For $N=2$ and $\bar{g}=c_i=1$\footnote{Which, by the way, is only possible 
for two coupled oscillators if $x_1=\frac{1}{4}(1\mp\sqrt{1+8\sqrt{14}}),
x_2=\frac{1}{2}-x_1,g_1=\frac{1}{2}\pm\frac{28175}{110608}\sqrt{1+8\sqrt{14}},
g_2=1-g_1$.} leads to the following solutions $E=-9.91107, -1.51155$ and
$E= 0.129506 \pm i 0.435961, 0.537995\pm i 1.32394$, which then corresponds 
to two bound 
states and two pairs of oscillatory states, the latter of which 
essentially oscillates around the ground state of the single harmonic 
oscillator.

\section{The Gaussian-Damped Anharmonic Oscillator}
The previous computations seem to suggest that the Sturmian method is best 
suited for potentials which are well behaved at infinity such as the Coulomb
potential, but very slow converging for potential diverging as $x\rightarrow
\infty$ such as the anharmonic oscillator. To test this hypothesis, we will
now briefly consider a toy model, the anharmonic oscillator damped by a 
Gaussian $V'=\alpha x^k e^{-x^2}$ where $k=3,4$.\footnote{One could also
consider simply the exponentially damped oscillator $V'=x^k e^{-|x|}$, but due
to the presence of the absolute value the matrix elements become too
complicated.} Including only the first
five Sturmians we get the following matrix elements, $\tilde{W}_{\bf 
nn'}^{(k)}$ of $V', k=3,4$
\beqa
	\tilde{W}^{(3)} &=& E^{1/2}\left(\begin{array}{ccc}
	0 & \frac{27}{2(3+4E)^{5/2}} & 0 \\
	\frac{27}{2(3+4E)^{5/2}} & 0 &-\frac{225(15-22E)}{4\sqrt{\frac{1}{6}+
	\frac{4E}{45}}(15+8E)^3} \\
	0 & -\frac{225(15-22E)}{4\sqrt{\frac{1}{6}+\frac{4E}{45}}(15+8E)^3} & 
	0 \\ 
	-\sqrt{\frac{3}{2}}\frac{343(3+2E)}{2(7+8E)^{7/2}} & 0 &
	\frac{1225(525-940E+316E^2)\sqrt{21}}{4\sqrt{1+\frac{12}{35}E}
	(35+12E)^4} \\ 
	0 & \frac{81(243-324 E-52E^2)}{4\sqrt{2+\frac{8}{9}E}(9+4E)^4} &
	0 
	\end{array}\right.\nonumber\\
	&&\qquad\left.\begin{array}{cc}
		-\sqrt{\frac{3}{2}}\frac{343(3+2E)}{2(7+8E)^{7/2}} & 0\\
	 	0 & \frac{81(243-324 E-52E^2)}{4\sqrt{2+
			\frac{8}{9}E}(9+4E)^4}\\
		\frac{1225(525-940E+316E^2)\sqrt{21}}{4\sqrt{1+\frac{12}{35}E}
			(35+12E)^4} & 0\\
		0 & -\frac{250047(107163-207522E+94356E^2-13816E^3)}
			{8(63+16E)^{11/2}}\\
		 -\frac{250047(107163-207522E+94356E^2-13816E^3)}
		{8(63+16E)^{11/2}} & 0
	\end{array}\right)\nonumber\\
	&&\\
	\tilde{W}^{(4)} &=& \left(\begin{array}{ccc}
	\frac{3}{4(1+2E)^{5/2}} & 0 & -\frac{75(5-4E)}{8\sqrt{\frac{1}{2}
	+\frac{3E}{5}}(5+6E)^3}\\
	0 & \frac{5}{2(1+2E/3)^{7/2}} & 0\\
	 -\frac{75(5-4E)}{8\sqrt{\frac{1}{2}+\frac{3E}{5}}(5+6E)^3} & 0 &
	\frac{75(25-80E+104E^2)}{8\sqrt{1+\frac{2}{5}E}(5+4E)^4}\\
	0 & -\frac{19845 E(21-4E)}{4\sqrt{\frac{1}{14}+\frac{5}{147}E}(21+10
	E)^4}\\
	\frac{243(243-160E^2)}{16\sqrt{\frac{3}{2}+\frac{5E}{3}}(9+10E)^4}
	& 0 &-\frac{2025(273375-899100E+1219680E^2-92288E^3)}{16\sqrt{\frac{1}
	{3}+\frac{14E}{135}}(45+14E)^5}
	\end{array}\right.\nonumber\\
	&&\qquad\left.\begin{array}{cc}
	0 & \frac{243(243-160E^2)}{16\sqrt{\frac{3}{2}+\frac{5E}{3}}(9+10E)^4}
	\\
	-\frac{19845 E(21-4E)}{4\sqrt{\frac{1}{14}+\frac{5}{147}E}(21+10
	E)^4} & 0\\
	0 & -\frac{2025(273375-899100E+1219680E^2-92288E^3)}{16\sqrt{\frac{1}
	{3}+\frac{14E}{135}}(45+14E)^5}\\
	\frac{735 E(147-112E+40E^2)}{4Sqrt{1+\frac{2E}{7}}(7+2E)^5} &0\\
	0 & \frac{729(19683-69984E+106272E^2-32256E^3+5248E^4)}{32 
	(9+2E)^{13/2}}
	\end{array}\right)
\eeqa
which leads to the ground state energies shown in table 4 below.
\noindent We notice the improved convergence properties supporting our claim
that it was the non-convergence of the matrix elements of the undamped
anharmonic oscillator that was the cause for the non-convergence of the
ground state energies in the Sturmian method.

\section{Conclusion}
We have seen that the powerful technique of Sturmians functions developed
for Coulomb-like potentials can be extended to harmonic and anharmonic 
oscillators, where it furthermore can be seen that the technique is highly
non-perturbative, but the divergence of the potential (as $x\rightarrow\pm
\infty$) leads to a non-convergence of the Sturmian approximation, in contrast
to the Coulomb case, where we have very rapid convergence. 
It turned out, however, that already with $N=1,2$ Sturmians the
correct order of magnitude for the energies of even the excited states
could be obtained. Thus indicating that the problem with convergence is
perhaps not so serious after all, if one merely wants to find the order of
magnitude. For higher precision, one should probably utilise a hybrid method, 
using the first few Sturmians to get the correct order of magnitude and then
some variational approach, say, to get the required precision.\\
We also saw how to treat time-dependent problems, where
once again the Sturmian properties lead to some important simplifications.
Finally, we considered a bath of coupled or uncoupled oscillators which
could be transformed into a single anharmonic oscillator problem. This is
contrary to what one does for the Coulomb potential, where Fourier
transform techniques are used in stead. Furthermore, when using a modified
(or regularised) potential, convergent at infinity but with the same behaviour
for $x$ not too large, we did get rapid convergence, especially for the $x^3$
case, whereas the $x^4$ case had slightly slower convergence.\\
All of this seems to suggest that the Sturmian techniques have a very wide 
range of applicability covering basically all important potentials known in
atomic physics or quantum chemistry, but one should be very careful when
using potentials which are not well behaved at infinity. 
Given the generality of the approach, as
outlined in the introduction, this procedure should also be extendible to
problems in quantum field theory using a functional Schr\"{o}dinger picture
and to problems in quantum kinetic theory in phase space using Wigner 
functions for instance.

\subsection*{Acknowledgements}
I thank John Avery for introducing me to Sturmians in the first place, and
for stimulating discussion during the writing of this paper.

\appendix
\section{Certain Results Concerning Hermite Polynomials}
In this appendix we prove a few results concerning Hermite polynomials. The
generating function is known to be
\beq
	e^{-s^2+sx} = \sum_{n=0}^\infty \frac{s^n}{n!}H_n(x)
\eeq
from this it is straightforward to deduce the standard orthonormality relation
for the Hermite polynomials and the harmonic oscillator wave functions. We will
also use it to derive some other useful results.\\
First we need the Fourier transform of a harmonic oscillator Sturmian. Thus
we want to compute the following integral
\beqs
	\int_{-\infty}^\infty e^{ikx}H_n(\alpha x)e^{-\frac{1}{2}\beta x^2} dx
\eeqs
The corresponding integral with the generating function is merely a Gaussian
integral and can be readily computed
\beqa
	\int e^{-s^2+\alpha sx-\frac{1}{2}\beta x^2 +ikx} dx &=&
	e^{-s^2}\sqrt{\frac{2\pi}{\beta}}e^{\frac{1}{2\beta}(\alpha s+ik)^2}\\
	&=& \sum_{n=0}^\infty \frac{s^n}{n!}\int H_n(\alpha x) e^{-\frac{1}{2}
	\beta x^2+ikx}dx
\eeqa
from which we get, by Taylor expansion, the desired result:
\beq
	\int H_n(\alpha x) e^{-\frac{1}{2}\beta x^2+ikx} dx = \sqrt{\frac{2\pi}
	{\beta}}\left(1+\frac{\alpha^2}{2\beta}\right)^{n/2} H_n(ik
	\frac{\alpha}{\beta}(1+\alpha^2/(2\beta))^{-1/2}) e^{\frac{1}{2}
	\beta^{-1}k^2}
\eeq
Actually, this formula is a little more general than we need. For harmonic
oscillator Sturmians it turns out that $\alpha=\beta$ ($=(\beta_{\bf n} 
m)^{1/4}$) which leads to a slight simplification, resulting in the formula
given in the text.\\
Another expression we need is the matrix element of $x^\gamma$ for $\gamma$
some positive integer $\geq 2$, i.e., we need to compute
\beq
	I_{nm}(\alpha,\beta,\gamma,\delta) := \int H_n(\alpha x) H_m(\beta x)
	x^\gamma e^{-\frac{1}{2}\delta x^2} dx
\eeq
Again, in terms of the generating function, the integral we need to compute
is quite simply
\beqa
	\int e^{-2s^2+2(\alpha+\beta)sx-\frac{1}{2}\delta x^2} x^\gamma dx
	&=&  e^{-2s^2} 2^{\gamma/2} \delta^{-1-\gamma/2}\Gamma(1+
	\frac{\gamma}{2})\times\nonumber\\
	&&\left(2(\alpha+\beta)s(1-(-1)^\gamma)~
	_1F_1(1+\frac{\gamma}{2};
	\frac{3}{2};2\frac{(\alpha+\beta)^2s^2}{\delta})+\right.\nonumber\\
	&&\left.(1+(-1)^\gamma)\sqrt{\frac{\delta\pi}{2}}e^{2\frac{(\alpha+
	\beta)^2s^2}{\delta}}L_{\frac{\gamma}{2}}^{-\frac{1}{2}}
	(-2\frac{(\alpha+\beta)^2s^2}
	{\delta})\right)\\
	&\equiv &{\cal I}(\alpha,\beta,\gamma,\delta)
\eeqa
which is valid even for non-integer $\gamma$. Here $_1F_1$ is a hypergeometric
function and $L_a^b$ is an associated Laguerre polynomial. The first few of 
these are
\beqas
	L_1^{-1/2} (x) &=& \frac{1}{2}-x\\
	L_2^{-1/2} (x) &=& \frac{1}{8}(3-12x+4x^2)\\
	L_3^{-1/2} (x) &=& \frac{1}{48}(15-90x+60x^2-8x^3)\\
	_1F_1(\frac{3}{2};\frac{3}{2};x) &=& e^x\\
	_1F_1(\frac{5}{2};\frac{3}{2};x) &=& \frac{1}{3}e^x(3+2x)\\
	_1F_1(\frac{7}{2};\frac{3}{2};x) &=& \frac{1}{15}e^x(15+20x+4x^2)
\eeqas
For the two cases
of interest to us, $\gamma$ an even or odd positive integer, we get a simpler
relation since one of the two terms on the right hand side will vanish.\\
For $\gamma=2k$ we arrive at
\beqs
	{\cal I}(\alpha,\beta,2k,\delta)=2^{k+1} e^{-2s^2}\delta^{-1-k} k!
	\sqrt{\frac{1}{2}\pi\delta} e^{2\frac{(\alpha+\beta)^2s^2}{\delta}}
	L_k^{-\frac{1}{2}}(-2\frac{(\alpha+\beta)^2s^2}{\delta})
\eeqs
whereas we for $\gamma=2k+1$ find
\beqs
	{\cal I}(\alpha,\beta,2k+1,\delta)=2^{k+5/2} e^{-2s^2}\delta^{-k-3/2}
	\Gamma(k+\frac{3}{2}) s (\alpha+
	\beta)~_1F_1(k+\frac{3}{2};\frac{3}{2};2\frac{(\alpha+\beta)^2s^2}
	{\delta})
\eeqs
Taylor expanding in $s$, we get
\beq
	{\cal I}(\alpha,\beta,\gamma,\delta) \equiv \sum_{n=0}^\infty
	{\cal I}_n(\alpha,\beta,\gamma,\delta) \frac{s^n}{n!}= 
	\sum_{n,m=0}^\infty 
	\frac{s^{n+m}}{n!m!}I_{nm}(\alpha,\beta,\gamma,\delta)
\eeq
From this we can obtain relationships between the matrix elements $W_{\bf nn'}
^{(k)}$ for different $k$'s, since $I_{nm}\propto W_{\bf nm}^{(k)}$ for
$\alpha=\beta_{\bf n}^{1/4}, \beta=\beta_{\bf m}^{1/4}$. For instance
\beqas
	I_{00} &=& {\cal I}_0\\
	I_{01}+I_{10} &=& {\cal I}_1\\
	\frac{1}{2}I_{02}+\frac{1}{2}I_{20}+I_{11} &=& \frac{1}{2}{\cal I}_2\\
	\frac{1}{6}(I_{03}+I_{30})+\frac{1}{2}(I_{21}+I_{12}) &=& \frac{1}{6}
	{\cal I}_3
\eeqas
such rules (essentially following from the recursion relation for the Hermite
polynomials) can be used to simplify the computation of matrix elements.\\
One should also note that, since the above is valid even for $\gamma$ not a 
positive integer, we can use it to obtain the matrix elements of the
Coulomb potential between harmonic oscillator Sturmians.

\newpage
\begin{table}[htb]
\center
\begin{tabular}{|c|l||c|l|}\hline
	$n$ & $N_n$ & $n$ & $N_n$\\ \hline
	0  & $\frac{1}{8\sqrt{2}}E^{-3/2}$ & 5 
		& $\frac{121}{8}\sqrt{\frac{11}{2}} E^{-3/2}$\\
	1  & $\frac{9}{8}\sqrt{\frac{3}{2}} E^{-3/2}$ & 6 & 
		$\frac{169}{8}\sqrt{\frac{13}{2}} E^{-3/2}$\\
	2  & $\frac{25}{8}\sqrt{\frac{5}{2}}E^{-3/2}$ & 7 
		& $\frac{225}{8}\sqrt{\frac{15}{2}}E^{-3/2}$\\
	3  & $\frac{49}{8}\sqrt{\frac{7}{2}}E^{-3/2}$ & 
		8 & $\frac{289}{8}\sqrt{\frac{17}{2}}E^{-3/2}$\\
	4  & $\frac{243}{\sqrt{2}}E^{-3/2}$ & 9 
		& $\frac{361}{8}\sqrt{\frac{19}{2}}E^{-3/2}$\\ \hline
\end{tabular}
\caption{Normalisation constants for harmonic oscillator Sturmians.}
\end{table}

\begin{table}[htb]
\center
\begin{tabular}{|c|l|l|}\hline
 $N$ & $E (x^3)$ & $E(x^4)$\\ \hline
 1 & 0.500000 & 0.562709\\ 
 2 & 0.014628 & 0.562709\\
 3 & 0.112767 & 0.562709\\
 4 & 0.351135 & 0.562544\\
 5 & 0.102981 & 0.562516\\ 
10 & 1.27012 & 0.533858\\ \hline 
\end{tabular}
\caption{The energies for the cubic and quartic anharmonic oscillator $V'=
\alpha x^3,
\alpha x^4$ with $\alpha=.1$ found by using 
only the first $N$ Sturmians. Only the ground state energies are shown.}
\end{table}

\begin{table}[htb]
\centering
\begin{tabular}{|c|r|r|r|r|r|}\hline
	\multicolumn{1}{|c|}{$n$} & \multicolumn{2}{c|}{$E$ (Sturmian)} & 
	\multicolumn{1}{c|}{$E$ (Bacus)} & 
	\multicolumn{2}{c|}{difference}\\ \cline{2-3}\cline{5-6} 
	& $N=1$ & $N=2$ & & $N=1$ & $N=2$\\ \hline
	0 & 1.07500 & 1.07500 & 1.06529 & -0.00971 & -0.00971\\
	1 & 3.37500 & 3.37500 & 3.30687 & -0.06813 & -0.06813\\
	2 & 5.97500 & 5.97500 & 5.74795 & -0.22705 & -0.22705\\
	3 & 8.87500 & 7.00152 & 8.35268 & -0.52232 & 1.35116\\
	4 & 12.0750 & 9.30093 & 11.09860 & -0.97640 & 1.79767\\ \hline
\end{tabular}
\caption{A comparison between the perturbed energy states for the quartic
anharmonic oscillator with $\alpha=.1$ found by using two Sturmians and the
high-precision results of Bacus et al., the second order perturbative result
for this case  ($m=\frac{1}{2},\omega=2$) is $E_0=0.4900$ for the ground 
state.}
\end{table} 

\begin{table}[htb]
\center
\begin{tabular}{|c||l|l|}\hline
	$N$ & $E(x^3)$ & $E(x^4)$\\ \hline
	1 & 0.500000 & 0.622877\\
	2 & 0.495852 & 0.622877\\
	3 & 0.491822 & 0.622878\\
	4 & 0.491282 & 0.622877\\
	5 & 0.491282 & 0.622878\\ \hline
\end{tabular}
\caption{The ground state energies for $V'=x^k e^{-x^2}, k=3,4$ computed
using $N$ Sturmians.}
\end{table}

\newpage

\begin{figure}[htb]
\caption{A plot of the first ten harmonic oscillator Sturmians showing how they
are scaled to all have essentially the same range.}
\end{figure}

\begin{figure}[htb]
\caption{The real (a) and imaginary (b) parts of the first time dependent
Sturmian for the damped (Caldirola-Kanai) oscillator, $c_n(t)\psi_0(x)$, in 
the range $t\in \left[0,10\right], x\in\left[0,5\right]$ and with $E=
\gamma=1$.}
\end{figure}

\begin{thebibliography}{99}
\bibitem{Sturm} H. Shull, P.O. L\"{o}wdin, J. Chem. Phys. {\bf 30} (1959) 617;
the method has a longer history and goes back to at least two very early 
papers: P. S. Epstein, Proc. Natl. Acad. Sci. (USA) {\bf 12} (1926) 637; 
B. Podolski, Proc. Natl. Acad. Sci. (USA) {\bf 14} (1928) 253
\bibitem{Rotenberg} M. Rotenberg, Adv. At. Mol. Phys. {\bf 6} (1970) 233; 
Ann. Phys. (NY) {\bf 19} (1962) 262.
\bibitem{Avery} J. Avery, {\em Hyperspherical Harmonics: Applications in 
Quantum Theory}, Kluwer Academic Publishers, Dordrecht (1989)
\bibitem{Avery2} J. Avery, D. Herschbach, Int. J. Quantum Chem. {\bf 41} 
(1992) 673.
\bibitem{Avery3} V. Aquilanti, J. Avery, J. Chem. Phys. Lett. {\bf 267} 
(1997) 1.
\bibitem{Avery4} J. Avery, J. Math. Chem. {\bf 21} (1997) 285; Adv. in Quant. 
Chem. (1998) (in press).
\bibitem{AA} J. Avery, F. Antonsen, J. Math. Chem. (1998) (in press).
\bibitem{BB} C. M. Bender, L. M. A. Bettencourt, Phys. Rev. Lett. {\bf 77}
(1996) 4114; Phys. Rev. D{\bf 54} (1996) 7710; C. M. Bender, T. T. Wu, Phys.
Rev. {\bf 184} (1969) 1231.
\bibitem{Kunihiro} T. Kunihiro, Phys. Rev. D{\bf 57} (1998) R2035.
\bibitem{Bacus} B. Bacus, Y. Meurice, A. Soemadi, J. Phys. A{\bf 28} (1995) 
L381.
\bibitem{others} W. Janke, H. Kleinert, Phys. Lett. A{\bf 206} (1995) 283;
Phys. Rev. Lett. {\bf 75} (1995) 2787; F. Vinette, J.
\v{C}i\v{z}ek, J. Math. Phys. {\bf 32} (1991) 3392; L. Salasnick, 
quant-ph/9803069; S. Biswas, K. Datta, R.
Sexena, P. Srivastava, V. Varma, J. Math. Phys. {\bf 14} (1973) 1190.
\bibitem{damped} P. Caldirola, Nouvo Cim. {\bf 18} (1941) 393; E. Kanai,
Progr. Theor. Phys. {\bf 3} (1998) 440; A. Mostafazadeh, Phys. Rev. A{\bf 55} 
(1997) 4084; J. Phys. A{\bf 31} (1998) 6495; S. S. Safanov, quant-ph/9802057,
to appear in Proc. VIII Int. Conf. on Sym. Meth. in Phys., Dubna 1997.
\end{thebibliography}
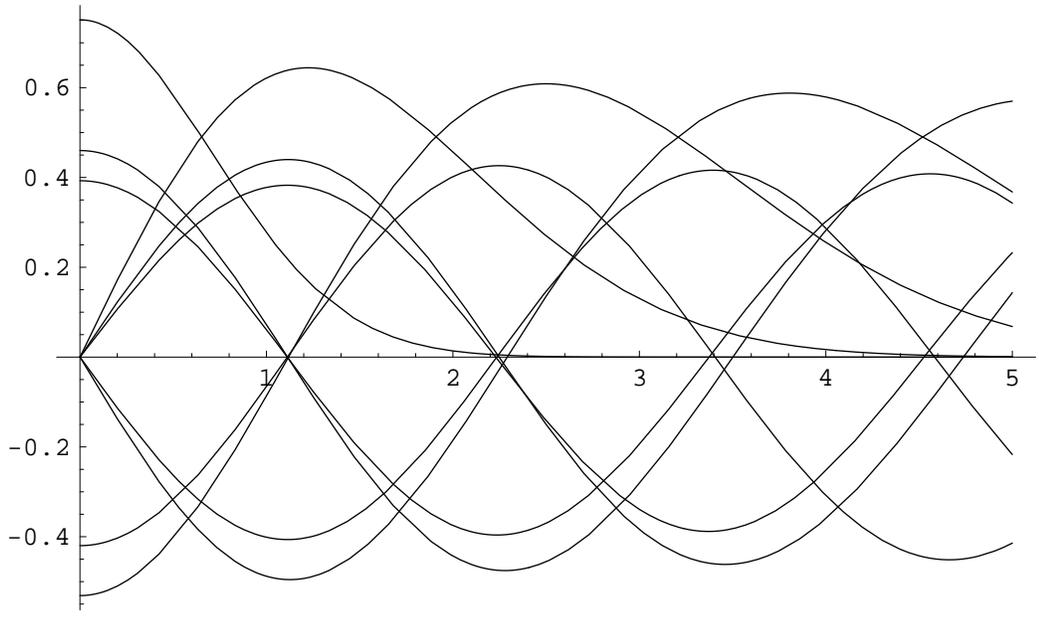
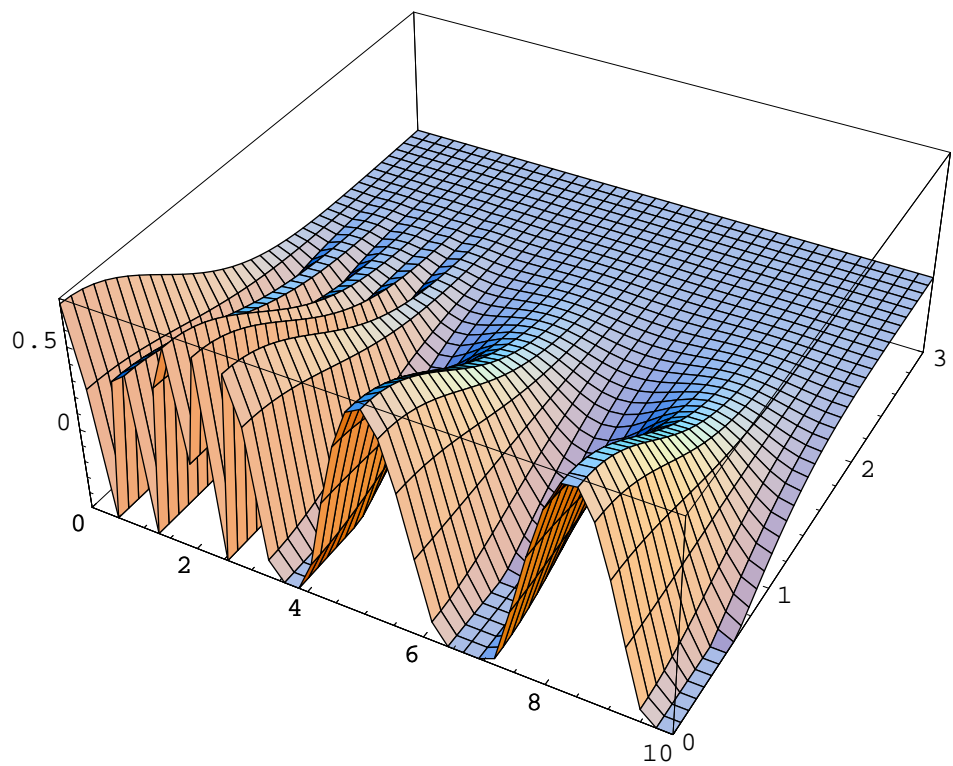
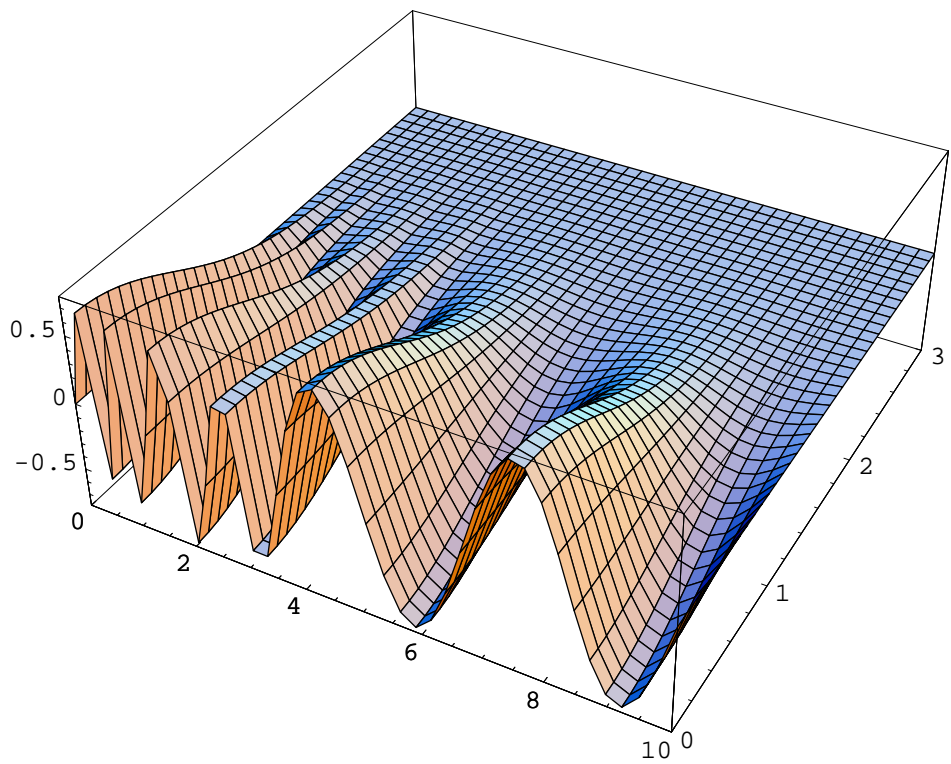
\end{document}